\documentclass[useAMS,usenatbib,onecolumn]{mn2e}

\usepackage{epsfig}
\usepackage{graphicx}
\usepackage{ulem}
\usepackage{color}

\DeclareMathAlphabet{\mathpzc}{OT1}{pzc}{m}{it}

\newcommand\smin{s_{\rm min}}
\newcommand\smax{s_{\rm max}}

\def\beq{\begin{equation}}
\def\eeq{\end{equation}}

\def\actaa{Acta. Astronom.} %
\def\aj{AJ}%
\def\apj{ApJ}%
\def\apjl{ApJ}%
\def\aap{A\&A}%
\def\mnras{MNRAS}%
\def\prd{Phys.~Rev.~D}%
\def\nat{Nature}%
\title[A New Photometric Model of the Bar]
      {A New Photometric Model of the Galactic Bar using Red Clump Giants}
\author[Cao, Mao, Nataf, Rattenbury \& Gould] {Liang Cao$^{1}$, Shude Mao$^{1, 2}$,
David Nataf$^{3,4}$, Nicholas J. Rattenbury$^2$, Andrew Gould$^4$  \\
$^{1}$ National Astronomical Observatories, 20A Datun Road, Chinese Academy of
Sciences, Beijing, 100012, China \\
$^{2}$ Jodrell Bank Centre for Astrophysics, The University of
Manchester, Alan Turing Building, Manchester M13 9PL, UK \\
$^{3}$ Research School of Astronomy and Astrophysics, Australian National University, Canberra, ACT 2611, Australia\\
$^{4}$ Department of Astronomy, The Ohio State University, 140 West 18th Ave., Columbus, OH 43210, USA
}
\begin{document}
\include{journaldefs}
\date{Accepted ...... Received ...... ; in original form......   }

\pagerange{\pageref{firstpage}--\pageref{lastpage}} \pubyear{2013}
\maketitle
\label{firstpage}

\begin{abstract}
We present a study of the luminosity density distribution of the Galactic bar using number counts of
red clump giants (RCGs) from the OGLE-III survey. The data were recently published by \citet{Nat12}
for 9019 fields towards the bulge and have $2.94\times 10^6$ RC stars over a viewing area of $90.25 \,\textrm{deg}^2$. The data include the number counts, mean distance modulus ($\mu$), dispersion in $\mu$ and full error matrix, from which we fit the data with several tri-axial parametric models. We use the Markov Chain Monte Carlo (MCMC) method to explore the parameter space and find that the best-fit model is the $E_3$ model, with the distance to the GC is 8.13 kpc, the ratio of semi-major and semi-minor bar axis scale lengths in the Galactic plane $x_{0},y_{0}$, and vertical bar scale length $z_0$, is $x_0:y_0:z_0 \approx 1.00:0.43:0.40$ (close to being prolate). The scale length of the stellar density profile along the bar's major axis is $\sim$ 0.67 kpc and has an angle of $29.4^\circ$, slightly larger than the value obtained from a similar study based on OGLE-II data. The number of estimated RC stars within the field of view is $2.78 \times 10^6$, which is systematically lower than the observed value. We subtract the smooth parametric model from the observed counts and find that the residuals are consistent with the presence of an X-shaped structure in the Galactic centre, the excess to the estimated mass content is $\sim 5.8\%$. We estimate the total mass of the bar is $\sim 1.8 \times 10^{10} M_\odot$. Our results can be used as a key ingredient to construct new density models of the Milky Way and will have implications on the predictions of the optical depth to gravitational microlensing and the patterns of hydrodynamical gas flow in the Milky Way.
\end{abstract}
\maketitle

\begin{keywords}
Galaxy - structure: Galaxy - stellar content - Galaxy: bulge - Galaxy: centre
\end{keywords}

\section{Introduction}
\label{sec:introduction}

Most spiral galaxies in the local Universe are barred. Galactic bars are often straight and their presence appears to be intimately related to star formation. Rings are often seen in these galaxies and they are likely related to resonances in a rotating barred potential. Thus barred galaxies offer exciting opportunities to understand gas and dynamical processes in spiral galaxies (see e.g. \citealt{SW93, ath12} for reviews). It is now generally accepted that the inner region of our own Galactic also hosts a tri-axial, bar-like structure. Observational evidence for a bar has arisen from several sources, such as the studies of gas kinematics (e.g. \citealt{1991MNRAS.252..210B}), surface brightness  (e.g. \citealt{1991ApJ...379..631B}), star counts (e.g. \citealt{1991Natur.353..140N,1994ApJ...429L..73S}) and microlensing (e.g. \citealt{1994AcA....44..165U}; see \citealt{2002ASPC..273...73G}  for a review).

Various tracers have been used to constrain stellar density models of the inner Galaxy. \citet{1995ApJ...445..716D} used the COBE-DIRBE multi-wavelength observations of the Galactic centre \citep{1994ApJ...425L..81W} to constrain several multi-parameter analytic bar models.  \citet{1997ApJ...477..163S} first used
red clump giants (RCGs) to constrain bar models (see also \citealt{2005MNRAS.358.1309B} and \citealt{2005ApJ...621L.105N} who traced the bulge RCG population in the infrared). \citet{Rat07} applied the method of \citet{1997ApJ...477..163S} to OGLE-II data and updated the bar parameters using the RCG population.

Unfortunately these studies leave many bar parameters uncertain. This may reflect the intrinsic complexities of the Galactic bar. For example, there has been a fair amount of discussion about the existence of a nuclear bar. \citet{2005ApJ...621L.105N} found a clear change of slope in the red clump giants around $|l|=4^\circ$, which was confirmed recently by the VVV survey (\citealt{Gon11}). This was interpreted as the existence of a nuclear bar. However, numerically simulated bars appear to be able to reproduce the changes in the slope without any nuclear structure \citep{GM12}. With more and more dynamical data emerging from several kinematic surveys (e.g., BRAVA, \citealt{How08}, ARGOS, \citealt{Fre13}, and APOGEE, \citealt{2012ApJ...755L..25N}), it appears increasingly important to construct a more accurate photometric model, which will serve as a key ingredient in self-consistent dynamical modelling (\citealt{Wan12, Lon13}).

The outline of this paper is as follows. In \S\ref{sec:data} we summarize the observational data used, as released by \citet{Nat12}; in \S\ref{sec:methodology} we describe the methodology used in this analysis; in \S\ref{sec:results}, we apply our methods to the OGLE-III data, and present the main results of this work. In \S\ref{sec:discussion} we discuss our results and compare them with earlier work.

\section{Data Sample}
\label{sec:data}

In this paper, we explore the OGLE-III observation data recently published by \citet{Nat12} who carefully studied the extinction and
the red clump giant population using both OGLE and 2MASS data. The RC is a prominent, well-populated, and localized feature of Galactic bulge CMDs (\citealt{1998AJ....115.1476T}; \citealt{1994ApJ...429L..73S}). RCGs are very good standard candles with $M_I=-0.12$ and an intrinsic dispersion of 0.09 mag in luminosity at the estimated age and metallicity distribution of the bulge stellar population\citep{Nat12}. A more detailed discussion of the effect of metallicity can be found in \S\ref{sec:metallicity}.

The final data of \citet{Nat12} contain 9019 sightlines, which are almost entirely within the range $-10^\circ <l < 10^\circ$ and $2^\circ<|b|<7^\circ$ and have a sky coverage of $90.25 \,\textrm{deg}^2$. The catalogue contains the total number counts, mean distance modulus, geometrical variance, i.e. square of the dispersion in distance modulus, and full error matrix for each sightline, which come from the extinction maps.

These new data address two shortcomings in the study by  \citet{Rat07}. First, the total area in \citet{Rat07} covered only about 11 square degrees, spread over a longitude range of $-6.8^{\circ} \leq l \leq 10.6^{\circ}$ degrees. Furthermore, the field latitudes are clustered around ($-4^\circ$) with only three fields at positive latitudes. Second, more seriously, \citet{Rat07} found that in order to match the luminosity of local and bulge red clump giants, an offset of 0.3 magnitude had to be artificially applied. The study by \citet{Nat12} addressed both issues. First, they resolve the ``offset" difficulty by a much more careful consideration of extinction and a new calibration in the intrinsic luminosity of the red clump giants. Second, the field covers larger than 90 square degrees, nearly one order of magnitude larger than that used in \citet{Rat07}. As a result, the differences between different models become much more drastic and the constraints on the model parameters are tighter. The data also allow us to explore the residuals between data and model much more convincingly, which we will exploit in some detail (see \S\ref{sec:results}).

\section{Methodology}
\label{sec:methodology}

In this section, we first describe the parametric tri-axial models we use to model the luminosity density model of the Galactic bar, then we outline the Markov Chain Monte Carlo method we use to find the best-fit parameters for each model. The results of our procedure will be presented in the next section.

\subsection{Parametric models}
We first re-examine a few models of bar density distribution proposed by \citet{1995ApJ...445..716D}:
\begin{equation}\label{eq:G1}
n_{\rm G_1} = N_0
\exp ({-r^2}/{2}),
\end{equation}
\begin{equation}\label{eq:E2}
n_{\rm E_2} = N_0 \exp(-r),
\end{equation}
\begin{equation}\label{eq:E3}
n_{\rm E_3} = N_0 K_0 (r_s),
\end{equation}
where $K_{0}$ is the modified Bessel function of the second kind and
\begin{eqnarray}\label{eq:r}
% \nonumber to remove numbering (before each equation)
  r &=& \left[\left(\frac{x}{x_{0}}\right)^{2} +
                  \left(\frac{y}{y_{0}}\right)^{2} +
                  \left(\frac{z}{z_{0}}\right)^{2}\right]^{1/2}\\
  r_{\rm s} &=& \left[\left[\left(\frac{x}{x_{0}}\right)^{2} +
                               \left(\frac{y}{y_{0}}\right)^{2}\right]^{2} +
                        \left(\frac{z}{z_{0}}\right)^{4} \right]^{1/4} .
\end{eqnarray}
These models are representative of a larger suite of parametric models studied by \citet{Rat07}.

We set the origin of the coordinate system at the Galactic Centre (GC), with the x-y plane aligned with the mid-plane of the Galaxy and the z-direction parallel to the direction of the Galactic poles. The x-direction is defined to be co-linear with the semi-major axis of the bar. The functions are rotated by an angle $\alpha$ around the z-axis.  $N_0$ is the number counts per $\rm kpc^3$, and the distance to the GC is another free parameter $R_0$. Thus we have six parameters to model the bar structure.

The bar can also be rotated by a tilt angle $\beta$ around the $y$ axis, corresponding to the Sun's position away from the mid-plane  of the Galaxy. Many previous studies found that the angle $\beta$ is close to zero \citep{Rat07,2012A&A...538A.106R}, nevertheless we include it as a seventh model parameter as a test. Furthermore, there is tentative evidence that the centroid of the bar may be offset from the centre of the Galaxy (as evidenced, for example, from the possible gamma-ray line emission offset from the centre, see \citealt{SF12}). Thus we also check the importance of two additional offset parameters, $\delta l$, $\delta b$. Finally, we also test the case for which both the tilt angle and the centre offset are included. We therefore fit the data using models in which all six, seven, eight or nine parameters are allowed to vary and compare the results with those found using the simplest six-parameter model.

With these models, we can calculate analytically the number counts, mean distance modulus and geometrical variance of each sightline as derived by \citet{Nat12}.

The number counts in each sightline $N_i$ is:
\begin{equation}
N_i = c_i \Delta \Omega \int_{\smin}^{\smax} n(s) s^2 ds,
\end{equation}
Where $\Delta \Omega$ is the solid angle subtended around each sightline, and $c_i$ is assumed to be a constant. We discuss the effects of metallicity on $c_i$ in \S\ref{sec:metallicity}. $\smin$ and $\smax$ are the minimum and maximum distances along the line of sight. We take $\smin=3$ kpc and $\smax=13$ kpc. We perform the integration over the range of $\sim R_0 \pm 5 {\rm kpc}$ as we do not expect the tri-axial bulge density structure to exceed these limits. The results are insensitive to the choice of integration limits since the bar density falls away rapidly and the contribution to the integral over the outer part of the bulge is negligible. Taking  $\sim R_0 \pm 3 {\rm kpc}$ instead, the total number count changes only by $0.2\%$ on average, and thus has the integration limit has little effect on our results.

The mean distance modulus of each sightline $\mu_i$ can be calculated as:
\begin{equation}
\mu_i = \int_{\smin}^{\smax} n(s) s^2  \mu ds\Big/ \int_{\smin}^{\smax} n(s) s^2 ds,
\end{equation}
where the distance modulus is $\mu = 5 \log(s/{\rm kpc})+10$.

The geometrical variance $\sigma_{\mu,i}$ can be written as,
\begin{equation}
\sigma^2_{\mu,i} = \left[\int_{\smin}^{\smax} n(s) s^2  \mu^2 ds \Big/ \int_{\smin}^{\smax} n(s) s^2 ds \right]- \mu_i ^2,
\end{equation}

We can evaluate the $\chi^2$ goodness-of-fit statistic using the number counts, mean distance modulus, geometrical variance, and their covariance matrix. The $\chi^2$ of the bar model can be written as,
\begin{equation}
\chi^2 = \sum_i (\textbf{X}_i^{\rm mod} - \textbf{X}_i^{\rm obs})^T \textbf{S}_i^{-1} (\textbf{X}_i^{\rm mod} - \textbf{X}_i^{\rm obs}).
\label{eq:chisq}
\end{equation}
$\textbf{X}_i^{\rm mod}$ comprises the number counts $(N_i)$, mean distance modulus $(\mu_i)$ and geometrical variance $(\sigma^2_{\mu, i})$  for the $i$-th sightline for the model under consideration, while $\textbf{X}_i^{\rm obs}$ comprises the observed values of these variables for the same sightline. $\textbf{S}_i$ is the covariance matrix of uncertainties. The data comprising $\textbf{X}_i^{\rm mod}$ are available from \citet{Nat12} and include 9019 lines of sight in total.

\subsection{Markov Chain Monte Carlo search for best-fit models}

For each of the parametric models, there are at least six parameters to constrain. We use Bayesian inference to provide a probabilistic distribution of tri-axial bar model parameters by combining prior information with the data. Information from the data and prior knowledge entirely determines the posterior probability distribution as follows:
\begin{equation}
\pi(x) \propto \mathcal{L}({\rm data}|x)\times {\rm prior}(x).
\end{equation}

We use Markov Chain Monte Carlo (MCMC) simulations to explore the parameter space and find the posterior distribution. MCMC can be implemented in a number of ways to explore the parameter space. In this work we use the standard random walk Metropolis-Hastings algorithm (\citealt{Metropolis}; \citealt{Hastings}). At any point $x_n$ in the chain we generate a new trial point $x_{\rm trial}$ by drawing parameter shifts from the symmetrical proposal distribution function $\phi(x-x_n)$. This algorithm is universal for any proposal distribution function. Here, we use a multi-dimensional Gaussian distribution in the vector of parameters.  We accept a candidate state $x_{n+1}$ with probability $\rm {min}\{1,\pi(x_{trial})/\pi(x_n)\}$, where $\pi(x)$ is the equilibrium distribution of the chain \citep{2002PhRvD..66j3511L}.

For each model, the chain is started at the corresponding best-fit model values of parameter vector as given in \citet{Rat07}. The model total star counts are normalised so that they reproduce the observed $N_0$ for each sightline. The MCMC simulation is then allowed to run, varying each of the six parameters in the range of [8.0, 8.5] (kpc), [0.0, 2.0] (kpc), [0.0, 1.0] (kpc), [0.0, 1.0] (kpc), [20.0, 40.0 ](deg), [0.0, 2.0$\times 10^7$](${\rm kpc}^{-3}$) for $R_0,x_0,y_0,z_0,\alpha,N_0$ respectively. Given the sampling strategy described above, we also use an adaptive step size to guarantee a sufficient sampling of the posterior, leading to convergence to the equilibrium posterior distribution in the chain after several thousand steps. We terminate the chains for each model according to its variance in the sample mean($\sigma^2_{\bar x}$) characterized by the dimensionless ratio $\sigma^2_{\bar x}/\sigma^2_0$ where $\sigma^2_0$ is the variance of the underlying distribution for convergence diagnostics \citep{2005MNRAS.356..925D}. We neglect the first half of the chain to avoid the influence of initial conditions, and then obtain the posterior distribution of the parameters.

\section{Results}
\label{sec:results}
\begin{table*}
\begin{tabular}{ccccccccccccccc}
  \hline
  % after \\: \hline or \cline{col1-col2} \cline{col3-col4} ...
  Model & $R_0$ (kpc) & $x_0$ (kpc) & $y_0$ (kpc) & $z_0$ (kpc) & $\alpha$ (deg)&$\delta l$ (deg)&$\delta b$ (deg)&$\beta$ (deg)&$\chi^2_1$& $\chi^2_2$&$\chi^2_3$&$\chi^2_4$&$\chi^2/d.o.f$ \\
  \hline
  $\rm G_1$ &8.19 & 1.18  & 0.49& 0.42 & 30.7 &&&&24160&33867&27850&7377&3.45\\
  $\rm E_2$ &8.16 & 0.66  & 0.27 &0.25 & 32.4 &&&&17094&31782&17842&3681&2.60\\
  $\rm E_3$ &8.13 & 0.67  & 0.29 &0.27 & 29.4 &&&&15675&31488&15989&1060&2.37\\
  $\rm G^{\prime}_1$ &8.22  & 1.18  & 0.49 &0.42 & 30.7 & & &-6.97&23897&33668&27880&7073&3.42\\
  $\rm E^{\prime}_2$ &8.18  & 0.67  & 0.27 &0.25 & 32.3 & & &-14.0&17126&31396&17851&3319&2.58\\
  $\rm E^{\prime}_3$ &8.13  & 0.67  & 0.29 &0.27 & 29.4 & & &0.11 &15675&31482&15989&1066&2.37\\
  $\rm G^{\prime \prime}_1$ &8.25  & 1.19  & 0.50 &0.42 & 30.9 &-0.38&-0.013&&23285&33950&27781&5922&3.36\\
  $\rm E^{\prime \prime}_2$ &8.20  & 0.69  & 0.27 &0.25 & 32.5 &-0.27&-0.024&&16965&31822&17768&2445&2.55\\
  $\rm E^{\prime \prime}_3$ &8.15  & 0.68  & 0.29 &0.27 & 29.5 &-0.10&-0.034&&15787&31488&15906&674&2.36\\
  $\rm G^{\prime \prime \prime}_1$ &8.25  & 1.19  & 0.50 &0.42 & 30.9 &-0.35&-0.016&-2.02&23319&33817&27800&5939&3.36\\
  $\rm E^{\prime \prime \prime}_2$ &8.20  & 0.67  & 0.27 &0.25 & 32.5 &-0.22&-0.031&-6.67&17058&31528&17774&2469&2.54\\
  $\rm E^{\prime \prime \prime}_3$ &8.15  & 0.68  & 0.29 &0.27 & 29.5 &-0.11&-0.034&0.59 &15797&31459&15904&686&2.36\\
  \hline
\end{tabular}
\caption{\label{tab:parameter} The best fit values of six, seven, eight and nine-parameter tri-axial galactic bar models. The $\chi^2$ contributed  from the number counts ($\chi^2_1$), mean distance modulus ($\chi^2_2$), geometrical variance ($\chi^2_3$) and the non-diagonal terms in the error matrix ($\chi^2_4$) are also listed.}
\end{table*}

For the simple bar models, we have six parameters $R_0, N_0, x_0, y_0, z_0$ and $\alpha$. We measure $2.94\times 10^6$ RC stars in the OGLE-III data over an area of $90.25 \,\textrm{deg}^2$. The total number of sightlines is 9019. For each model, we run the MCMC chains and obtain the posterior distribution of each parameter and its best fit value for each model.

The results of best fit parameters are presented in Table \ref{tab:parameter}.
Since $N_0$ does give the normalization for the total number of RC stars in the bulge, but it doesn't effect the spatial distribution of the model. Thus we include $N_0$ in the MCMC chain but do not give its value here. We separate $\chi^2$ contributions from the number counts ($\chi^2_1$), mean distance modulus ($\chi^2_2$), geomertical variance  ($\chi^2_3$) respectively. $\chi^2_4$ is the contribution from non-diagonal terms in the error matrix. Among these models, the $E_3$ model has the smallest $\chi^2/d.o.f$ values and the $G_1$ model has the largest $\chi^2/d.o.f$. This suggests that the $E_3$ model best fit the data, while the $G_1$ model fits the data worst.

\begin{figure}
  % Requires \usepackage{graphicx}
  \includegraphics{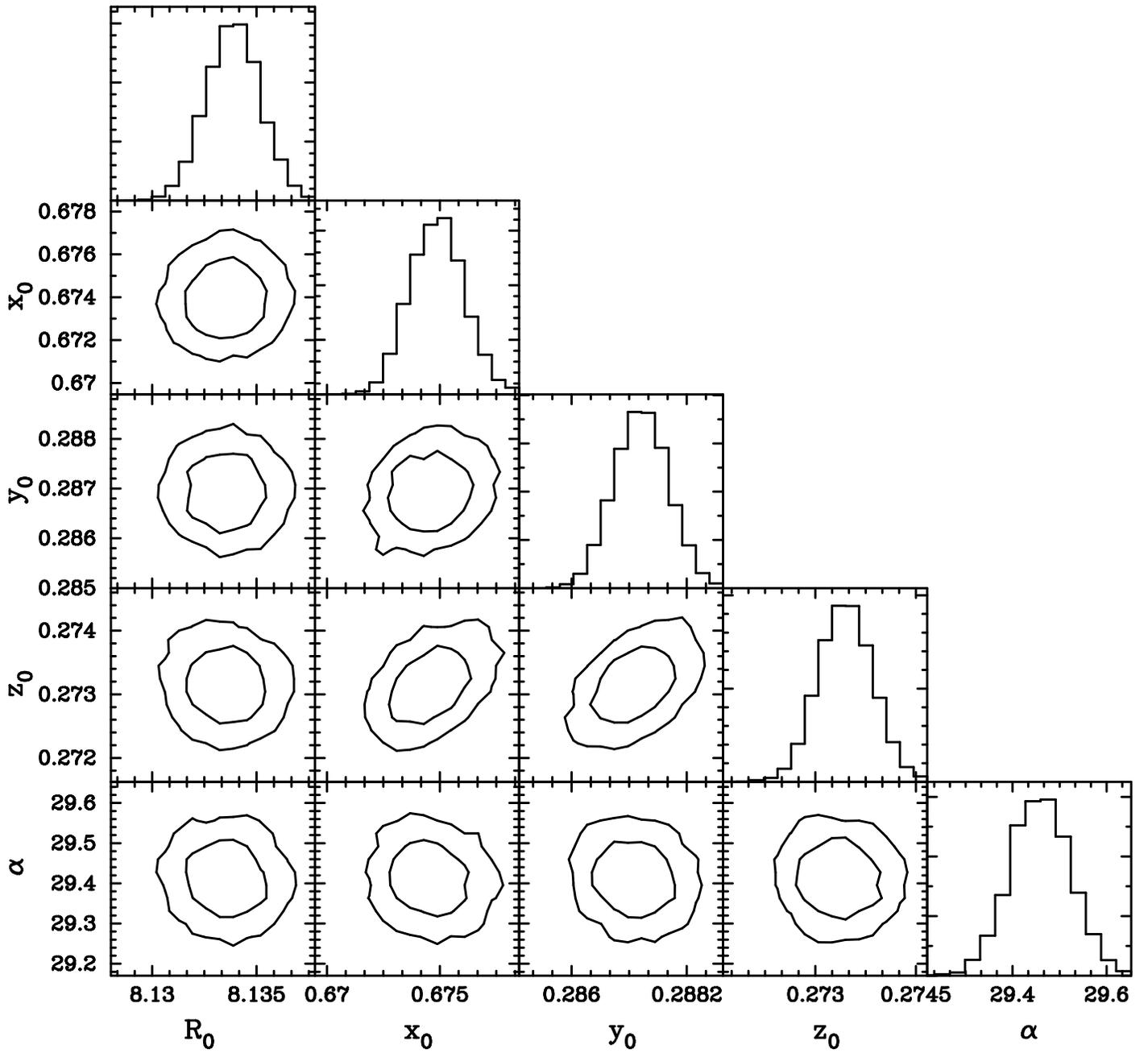}\\
  \caption{The marginalised probability density distributions (PDFs) and contours in the plane of $R_0$, $x_0$, $y_0$,$z_0$, $\alpha$ for the six-parameter $E_3$ model. All the lengths are in units of kpc and the angle $\alpha$ is measured in degrees. The inner and outer contours are respectively for the 1-$\sigma$ and 2-$\sigma$ confidence level regions (for two parameters simultaneously). }
  \label{pdf:E3}
\end{figure}

We present the marginalized posterior distribution of the parameters for the six-parameter $E_3$ model in Fig.~\ref{pdf:E3}. We can see that the probability distribution function (PDF) of each parameter is nearly Gaussian. We also show the degeneracy between each parameter in Fig.~\ref{pdf:E3}. The 1- and 2-$\sigma$ regions (i.e., 68\% and 95\% confidence levels) are shown in inner and outer contour lines respectively.

\begin{figure}
% Requires \usepackage{graphicx}
\includegraphics{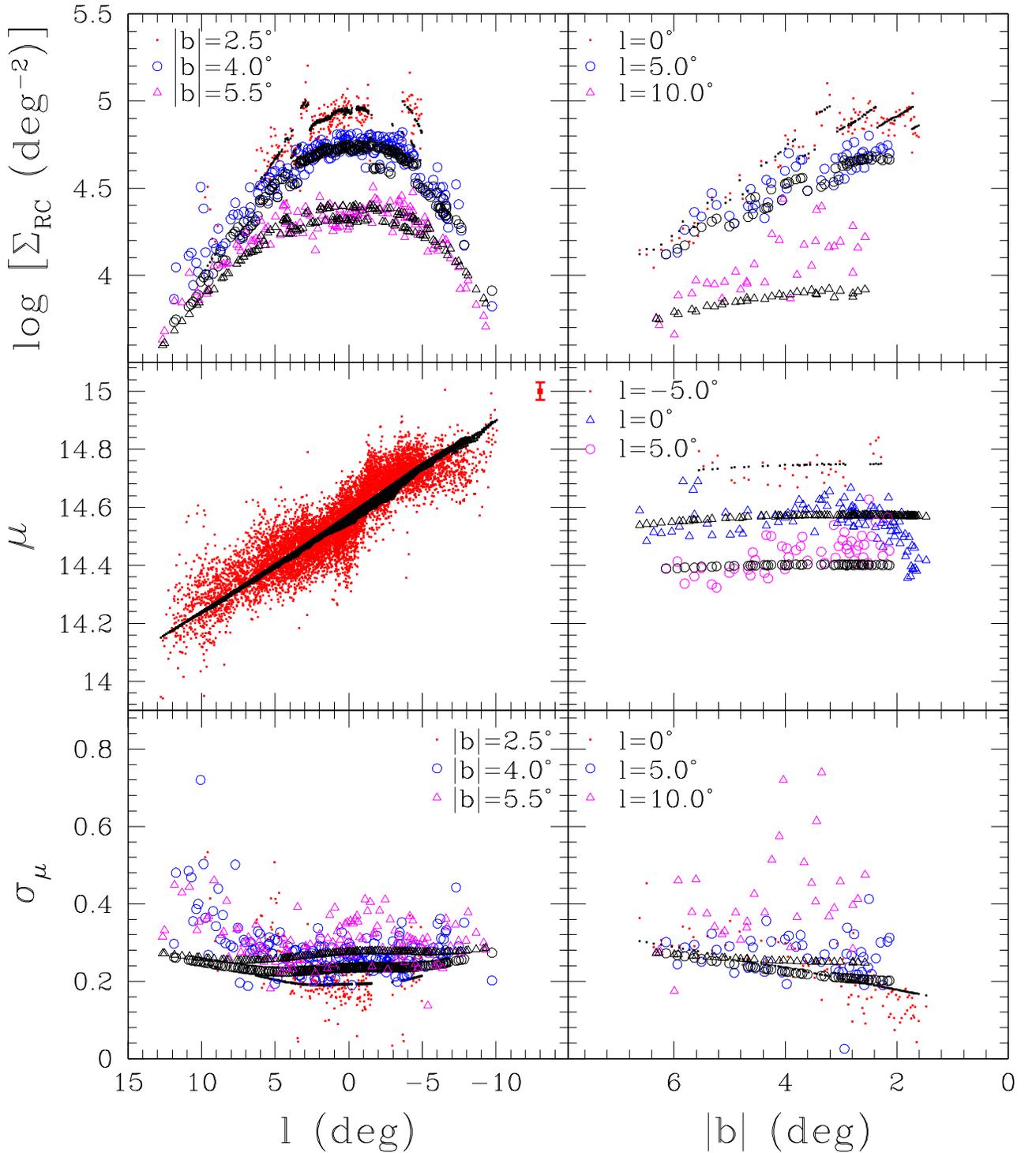}\\
\caption{The predicted surface density of stars $\Sigma_{RC}$, mean modulus $\mu$, and dispersion in modulus $\sigma_\mu$, vs. those directly from observation $E_3$ model. Top panels: the surface number density $\Sigma_{RC}$ vs. longitude (left) and latitude (right). Middle panels:  mean distance modulus $\mu$ vs. longitude (left) and latitude (right).  Bottom panels: dispersion in modulus $\sigma_\mu$  vs. longitude (left) and latitude (right). In all panels, black symbols denote model values while colour symbols denote observational data. The point with error bar at middle left panel is the median error on the measurement reported by \citet{Nat12}.}\label{fig:E3}
\end{figure}

We compare the model prediction quantities, e.g. the surface density of stars $\Sigma_{RC}$, mean distance modulus $\mu$, and geometrical variance $\sigma_{\mu}$, with that directly from observation for the model $E_3$. In the top left panel of Fig. \ref{fig:E3}, we plot $\Sigma_{RC}$ for three stripes at $|b|=2.5^\circ$, $4.0^\circ$, $5.5^\circ$ as observed in \citet{Nat12}. $\Sigma_{RC}$ is maximized near $l=0^\circ$ for all latitude stripes. In the same figure, we also plot the $E_3$ model predictions of $\Sigma_{RC}$. At small longitudes, the model predictions are consistent with the observations. However, at large longitudes, especially for the $|b|=5.5^\circ$ stripe, the model predictions are systematically lower than the observation. In the top right panel of Fig. \ref{fig:E3}, we plot $\Sigma_{RC}$ from both model predictions and observations at fixed longitudes of $l=0.0^\circ$, $5.0^\circ$ and $10.0^\circ$ respectively. It shows the expected result that number counts increase monotonically with decreased separation from the plane for each longitude. However, it again shows that at large longitude, especially for the $l=10.0^\circ$ stripe, the model predictions $\Sigma_{RC}$ are systematically lower than the observed values.

The middle left panel of Fig. \ref{fig:E3} shows the $E_3$ model (black symbols) and observed (red symbols) mean distance modulus $\mu$ against longitude. In the middle right panel, we plot $\mu$ at fixed longitudes of $l=-5.0^\circ$, $0^\circ$ and $5.0^\circ$. The bottom panels of Fig. \ref{fig:E3} is the same as the top panel, but for the dispersion in distance modulus $\sigma_\mu$. The dependence on longitude and latitude is reproduced, but the model prediction does not match the observation perfectly. In particular, for the $l=10^\circ$ stripe, the prediction is systematically lower than the observation.

Notice that negative values of $\sigma^2_\mu$ should not be a cause of concern. \citet{Nat12} estimated the dispersion due to geometric extension $\sigma^2_\mu$ as:
\begin{equation}
\sigma^2_\mu = \sigma^2 -\sigma_I^2 - \sigma_E^2
\end{equation}
where $\sigma^2$ is the measured brightness dispersion of the RC, $\sigma_I=0.09$ is the estimated intrinsic magnitude dispersion of RC, and  $\sigma^2_E$ is the dispersion due to differential extinction. Therefore, for sightlines where the measured brightness dispersion is small relative to the intrinsic dispersion and differential extinction, we can sometimes obtain negative variance in $\sigma^2_\mu$. We do not remove these sightlines with negative geometrical variance to avoid systematically biasing the final output to have large variance. However, in the bottom panels of Fig. \ref{fig:E3} the predicted $\sigma_\mu$ are well concentrated while the data have very wide variations.
It could be that the models are wrong, or it could be that the data
are wrong. To test for the latter, we take the intrinsic dispersion of the RC as
a free parameter instead of  $\sigma_I=0.09$ as quoted in \cite{Nat12}. The results are presented in Table \ref{tab:sigma}. We can see  different values of $\sigma_I$ are favored for different models: for the $G_1$ model, the best-fit $\sigma_I$ is preferred to be 0.11; for models $E_2$ and $E_3$,  $\sigma_I=0$ is preferred. The differences in the $\chi^2$ are formally important. However, since our models do not have $\chi^2/d.o.f$ close to one, which indicates systematic departures from our smooth ellipsoidal model (e.g., the X-shaped structure), we thus do not regard the inclusions of the additional parameters as very significant (i.e., the value of $\sigma_I$ is not well constrained by the spatial density models), and thus will not be discussed any further.

\begin{table*}
\begin{tabular}{ccccccccccccccc}
  \hline
  % after \\: \hline or \cline{col1-col2} \cline{col3-col4} ...
  Model & $R_0$ (kpc) & $x_0$ (kpc) & $y_0$ (kpc) & $z_0$ (kpc) & $\alpha$ (deg)&$\sigma_I$&$\chi^2_1$& $\chi^2_2$&$\chi^2_3$&$\chi^2_4$&$\chi^2/d.o.f$ \\
  \hline
  $\rm G_1$ &8.19 & 1.15  & 0.48& 0.42 & 31.9 &0.11&24267&33383&29546&6015&3.45\\
  $\rm E_2$ &8.16 & 0.69  & 0.29 &0.25 & 30.4 &0.00&16733&31878&19344&1430&2.57\\
  $\rm E_3$ &8.12 & 0.70  & 0.30 &0.28 & 27.3 &0.00&15469&31395&17967&-1474 &2.34\\
  \hline
\end{tabular}
\caption{\label{tab:sigma} The same as Table \ref{tab:parameter}, but taking the intrinsic dispersion of the RC $\sigma_I$ as
a free parameter.}
\end{table*}

\begin{figure}
\includegraphics{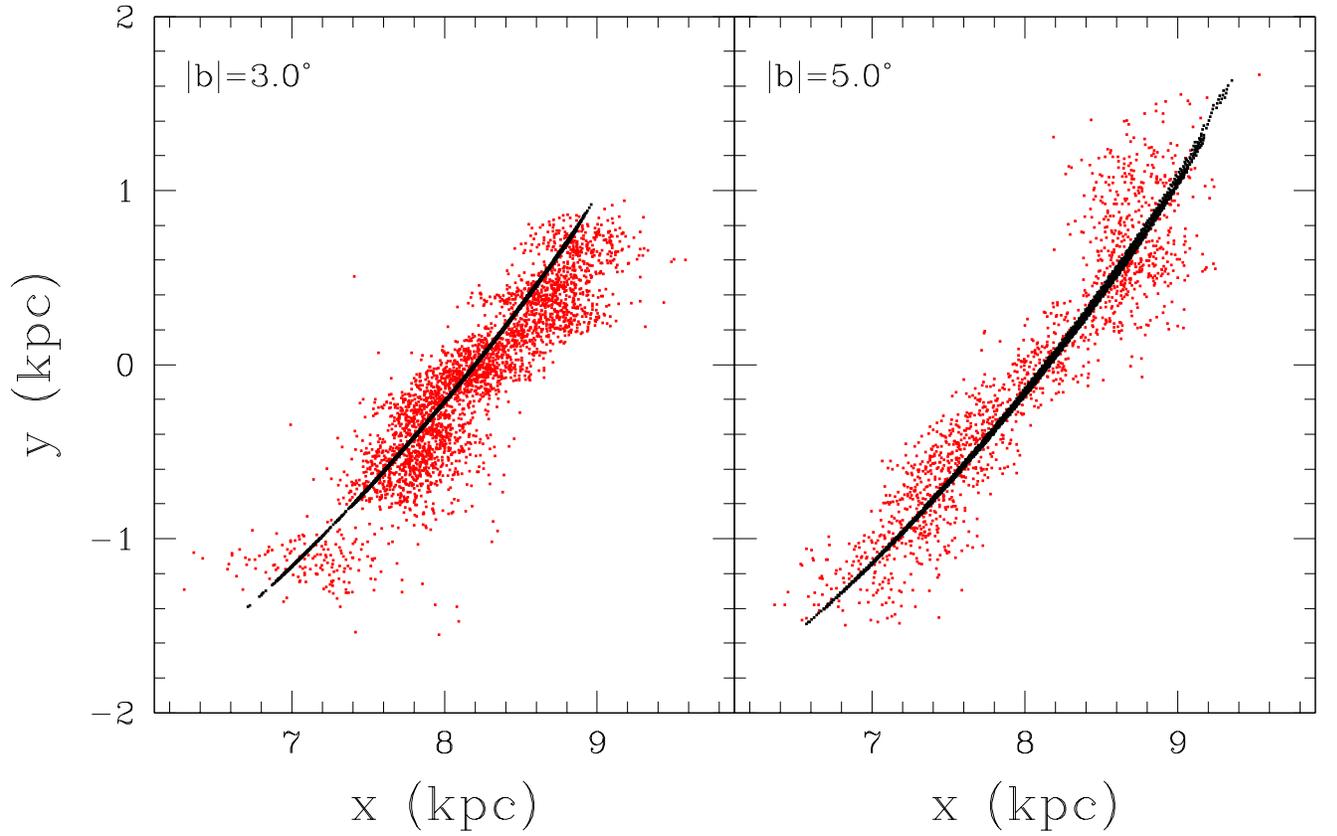}
\caption{The face-on view projection of the distances to RC centroids for the $E_3$ model. The red points show the projected location of each measured RC centroid, and black points show the model prediction. The left panel is for sightlines satisfying $2.5^\circ\leq |b| \leq 3.5^\circ$, while the right panel is for $4.5^\circ\leq |b| \leq 5.5^\circ$ sightlines.} \label{distance:E3}
\end{figure}

In Fig. \ref{distance:E3}, we plot the mean distance to RC centroids measured by \citet{Nat12} and give the $E_3$ model predictions in the same figure for comparison. The distances are projected onto a face-on view of the central region of the Galaxy. The left panel of Fig. \ref{distance:E3} shows data for sightlines with $2.5^\circ\leq |b| \leq 3.5^\circ$, while the right panel is for sightlines with $4.5^\circ\leq |b| \leq 5.5^\circ$. The model prediction is broadly consistent with the observations, but does not yield a very significant flattening at small positive longitudes as is seen in the data.

An interesting phenomenon, recently discovered, is that the RC of the bulge is observed to show two peaks
in distance at large separations from the plane \citep{Nat10, 2010ApJ...724.1491M, Sai11}, a result that has been
shown to be due to an X-shape structure by means of N-body models \citep{2012ApJ...757L...7L, 2012ApJ...756...22N}. The existence of the vertical X-shaped structure hints that the formation of the Milky Way bulge is shaped mainly by internal disk dynamical instabilities.

\begin{figure}
\centering\includegraphics{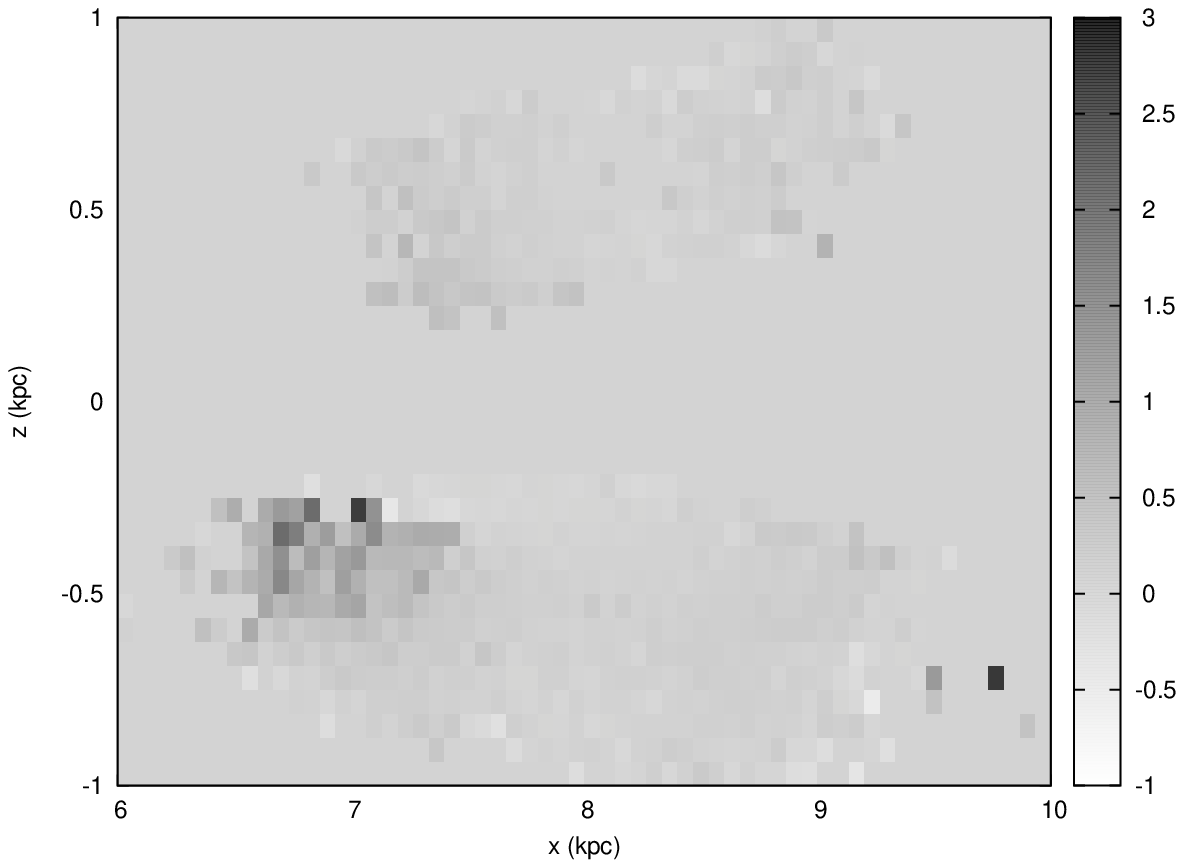}\\
\centering\includegraphics{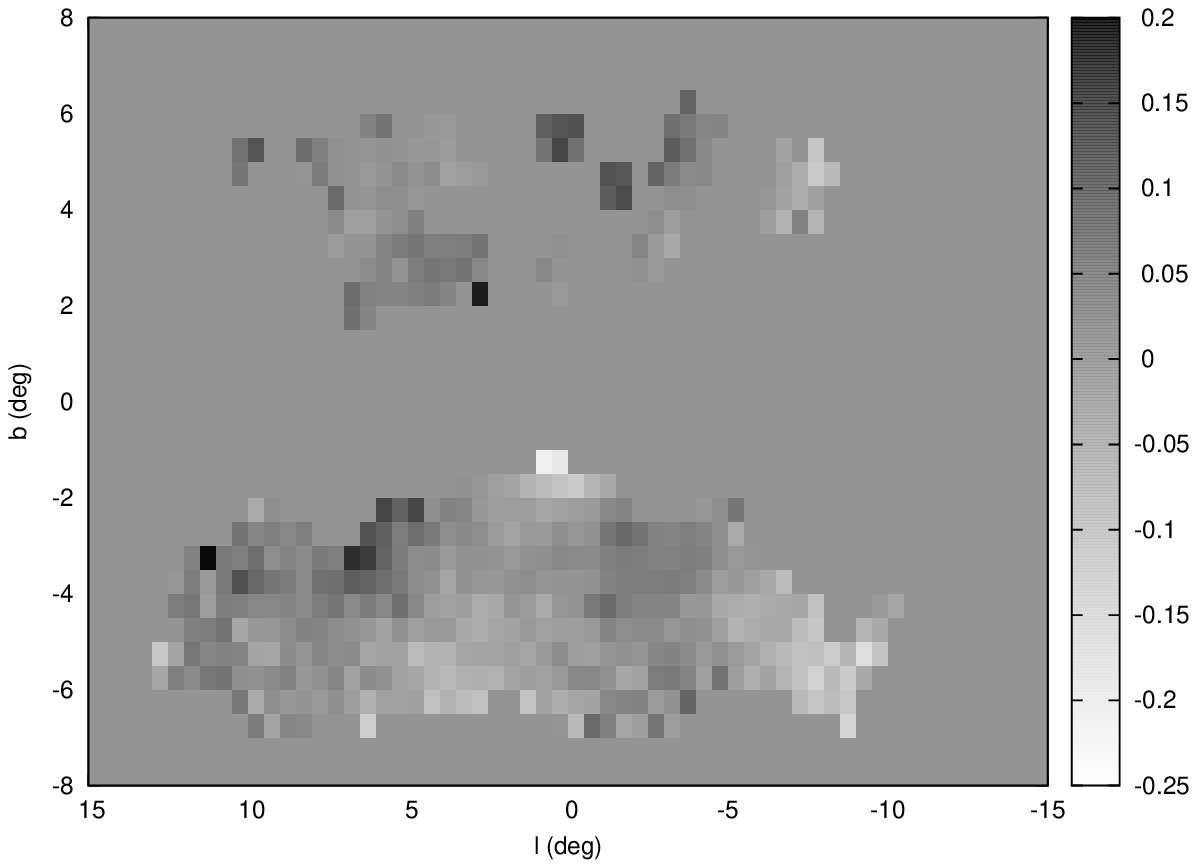}\\
\caption{The residual map of number counts  for the six-parameter $E_3$ model for edge-on (top) and side-on view (bottom) respectively. The residual of number counts is defined as the fractional difference in star counts between the model and the observed data, $(N_{\rm obs}-N_{\rm model})/N_{\rm model}$. The total excess fraction of number counts is 5.8\%.} \label{residual:E3}
\end{figure}

To verify the X-shaped fine structure, we calculate the excess fraction in star counts between the model and the observed data, $(N_{\rm obs}-N_{\rm model})/N_{\rm model}$, for each sightline. We then project the excess fraction to produce a 2-D residual map in the X-Z plane, shown in the top panel of Fig.~\ref{residual:E3} for the $E_3$ model. We can see a dim X-shaped structure extending from the x-axis $\sim$6 kpc to $\sim$10 kpc, with $|z|\leq 1$ kpc. We observe $2.94\times 10^6$ stars, while we find a predicted number of RCs for the bulge of $2.78\times 10^6$, implying an excess RC population of 5.8\%. The existence of an X-shape structure suggests that at the outer regions of the bulge, there is some fine structure in the Galactic bar. In the bottom panel of Fig.~\ref{residual:E3}, we project the excess fraction in the l-b plane and verify that the X-shape structure is also apparent in the side-on view.

As mentioned above, we also consider more complex models by introducing the tilt angle $\beta$, a centroid offset $(\delta l, \delta b)$, and the tilt angle and centroid offset together. To avoid confusion, these models are labelled as $G^\prime_1, E^\prime_2$ and $E^\prime_3$, $G^{\prime \prime}_1, E^{\prime \prime}_2$ and $E^{\prime \prime}_3$, $G^{\prime \prime \prime}_1, E^{\prime \prime \prime}_2$ and $E^{\prime \prime \prime}_3$. The fitting results are presented in Table~\ref{tab:parameter}. As we mentioned before, while the improvement of $\chi^2$ due to the additional parameters is formally significant, in light of the large $\chi^2/d.o.f$ for our best-fit model. we do not pursue these parameters any further. This is reasonbale since our sightlines do not cover the plane and very central regions which would presumably show the most prominent deviations from a tilt and non-offset model. Thus from now on $\beta$, $\delta l$ and $\delta b$ will all be fixed at zero.

\subsection{Disk contamination}

\begin{table*}
\begin{tabular}{ccccccccccccc}
  \hline
  % after \\: \hline or \cline{col1-col2} \cline{col3-col4} ...
  Model &Areas&Sightlines&$R_0$ (kpc) & $x_0$ (kpc) & $y_0$ (kpc) & $z_0$ (kpc) & $\alpha$ (deg) &$\chi^2_1$& $\chi^2_2$&$\chi^2_3$&$\chi^2_4$&$\chi^2/d.o.f$ \\
  \hline
  $G_1$&$|l|\leq4^\circ$&5995&8.19 & 1.26  & 0.42 & 0.40 & 29.2 &14683&23058&19067&1762&3.26\\
  $E_2$&$|l|\leq4^\circ$&5995&8.16 & 0.77  & 0.24 & 0.24 & 31.6 &11025&21375&12707&832&2.56\\
  $E_3$&$|l|\leq4^\circ$&5995&8.14 & 0.71  & 0.25 & 0.27 & 27.7 &9647&20829&10402&734&2.31\\
  $G_1$&$|l|\leq4^\circ$ and $|b|\leq4^\circ$&4405&8.19 & 1.08 & 0.39 & 0.37 & 27.8 &8182&18415&11109&280&2.88\\
  $E_2$&$|l|\leq4^\circ$ and $|b|\leq4^\circ$&4405&8.16 & 0.68 & 0.23 & 0.25 & 31.2 &7043&16903&8170&-63&2.47\\
  $E_3$&$|l|\leq4^\circ$ and $|b|\leq4^\circ$&4405&8.14 & 0.68 & 0.25 & 0.28 & 28.1 &6606&16385&8006&-40&2.34\\
  \hline
\end{tabular}
\caption{\label{tab:parameter_cut} The same as Table \ref{tab:parameter}, but removing the outer part of the bulge regions.}
\end{table*}

There is evidence that there is some structural change in the Galactic bulge inward of $l\leq 4^\circ$ \citep{GM12}. As mentioned above, the vertical X-shaped structure is assumed to be shaped mainly by internal disk dynamical instabilities and extend to large distance. Both features are a source of disk contamination which is difficult for our model to reproduce, especially in regions remote from the centre of the bulge. From Fig. \ref{fig:E3}, we can see that at larger longitude, the number counts and geometrical variance in observation also systematically deviate from the model prediction.

To see the effects of possible disk contamination, we simply exclude data from the outer part of the bulge. We adopt two cut criteria. Firstly, we remove  sightlines with $|l|>4^\circ$. Secondly, we retain only the sightlines whose longitude and latitude are both smaller than $4^\circ$. The value of 4 degrees is chosen because in this region, there is a slope change in the particle surface density, see, e.g., Fig. 2 of \citet{GM12}.

\begin{figure}
  % Requires \usepackage{graphicx}
  \includegraphics{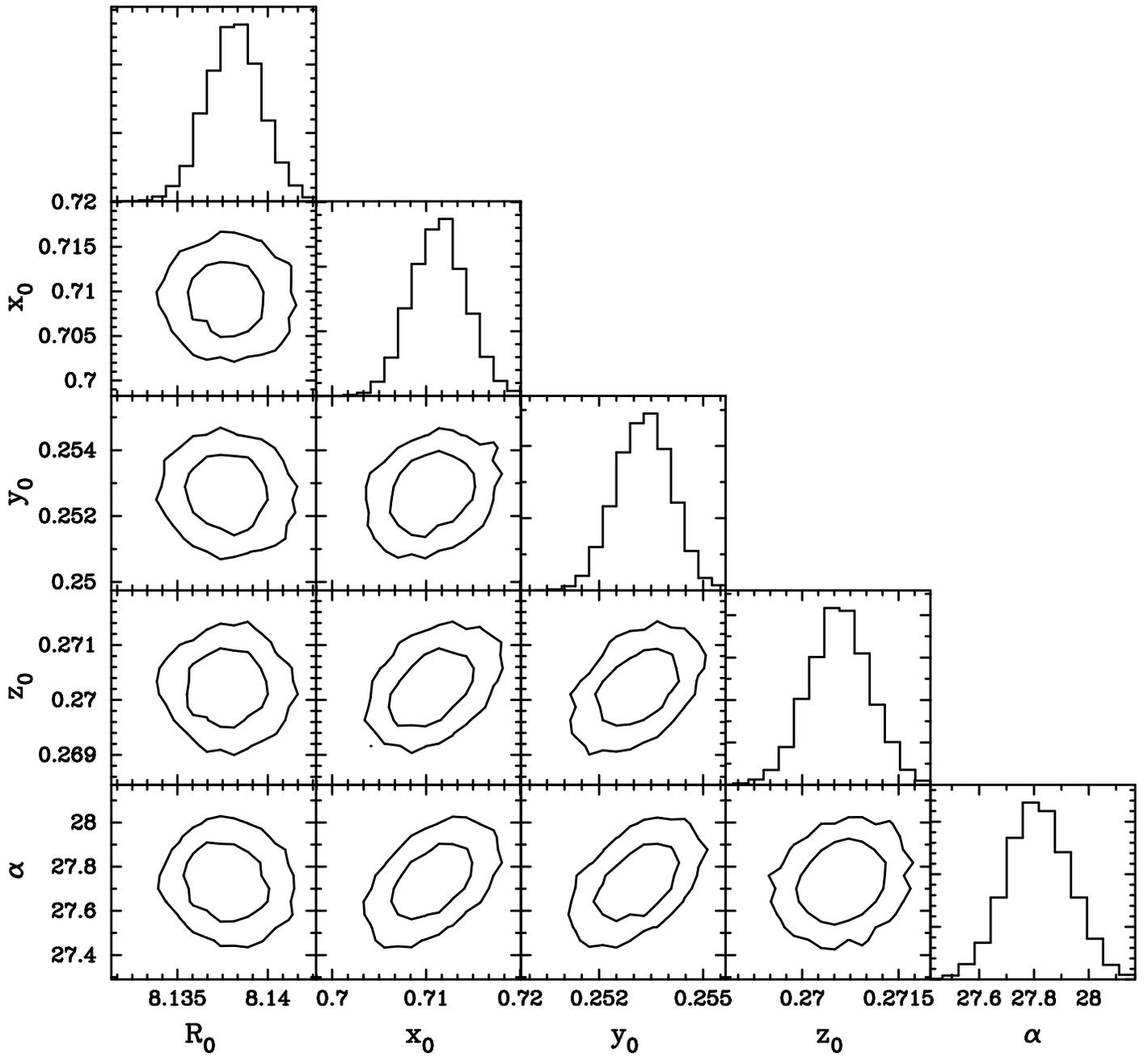}\\
  \caption{The same as Fig. \ref{pdf:E3}, but after excluding the $|l|>4^\circ$ areas.}
  \label{pdf:E3_lcut}
\end{figure}

\begin{figure}
  % Requires \usepackage{graphicx}
  \includegraphics{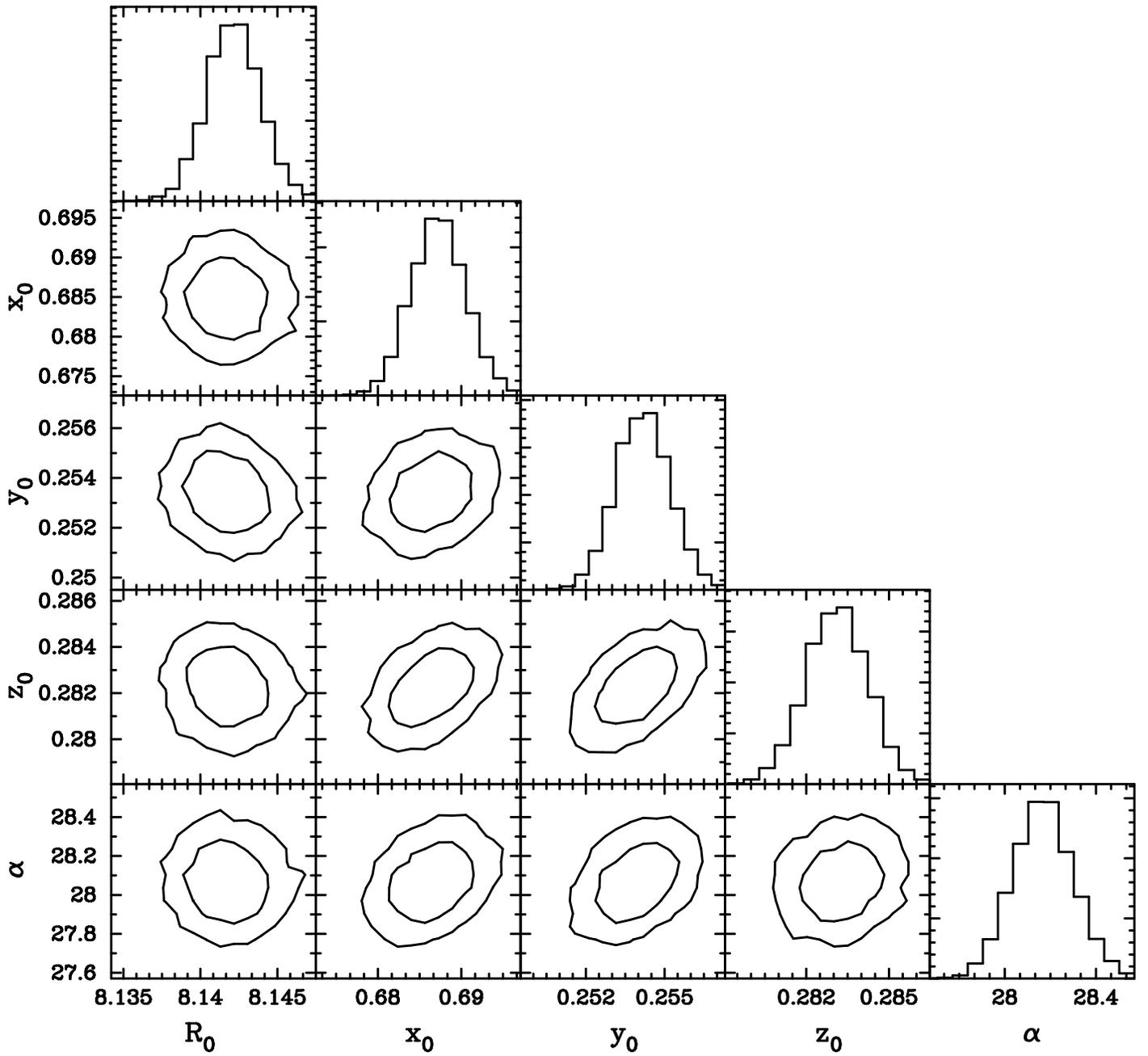}\\
  \caption{The same as Fig. \ref{pdf:E3}, but after excluding the $|l|>4^\circ$ and $|b|>4^\circ$ areas.}
  \label{pdf:E3_lbcut}
\end{figure}

We test the results of the models for the above two criteria (Table \ref{tab:parameter_cut}). Of course, after excluding the outer part, the resulting region is smaller with respect to the original size of the OGLE-III data. If we exclude the $|l|>4^\circ$ areas, there are 5995 sightlines remaining and the area is reduced to $40.67 \,\textrm{deg}^2$. Similarly, if we exclude the $|l|>4^\circ$ and $|b|>4^\circ$ areas, there are 4405 sightlines remaining in $17.81 \,\textrm{deg}^2$. Compared with the models for the whole dataset, excluding the outer part of the bulge results in better models in terms of
$\chi^2/d.o.f$ (see Table \ref{tab:parameter_cut}). We present in Figs. \ref{pdf:E3_lcut} and \ref{pdf:E3_lbcut} the PDF and contours of the parameters of $E_3$ when we exclude $|l|>4^\circ$ only and both $|l|>4^\circ$ and $|b|>4^\circ$ respectively. We again calculate the model-predicted RC population, and find the excess is reduced to 3.2\% for excluding fields with $|l|>4^\circ$ and 2.9\% for $|l|>4^\circ$ and $|b|>4^\circ$. These are not surprising since as we can see from Figs. \ref{fig:E3} and \ref{residual:E3} both the larger dispersion and excess lie in the outer region of the bugle. This suggests that we can obtain a more realistic inner bar structure after excluding the large longitude sightlines. The cuts on the data partly avoid the disk contamination and fine structure, which are presently very difficult to handle.

\subsection{Comparison with results from OGLE-II}

\citet{Rat07} used RCG stars from 44 bulge fields from the OGLE-II microlensing collaboration database to constrain analytical tri-axial models for the Galactic bar. They found the bar major axis is oriented at an angle of $24^\circ$ to $27^\circ$ to the Sun-Galactic centre line of sight. They also found the ratio of semi-major and semi-minor bar axis scale lengths in the Galactic plane $x_0$, $y_0$, and vertical bar scale length $z_0$, to be $x_0:y_0:z_0=1.00:0.35:0.26$.

As mentioned in the introduction, the data from OGLE-III as analyzed by \citet{Nat12} offer many larger sky coverage and better calibration in the RCG population. Both are helpful to refine the constraints on the bar parameters. We obtain a bar angle oriented at $29.4^\circ$, and the axis ratios are around $x_0:y_0:z_0=1.00:0.43:0.40$ with the scale length of the major axis being 0.67 kpc for the $E_3$ model. The scale length of the major axis is shorter than that found by \citet{Rat07} (1.00 kpc), while similar to \citet{1995ApJ...445..716D}(0.69 kpc). This suggests a bulge that is not as tri-axial as that found by \citet{Rat07}. The bar is also vertically more extended than the working model of \citet{2002ASPC..273...73G} with $x_0:y_0:z_0=1.00:0.40:0.30$. We notice that the scale lengths for different models are different, the Gaussian model has the largest scale length among all the models. An exponential falls off more slowly than a Gaussian; it has a longer tail. So to include the same number of stars within some physical scale, an exponential fit will have a smaller scale length than a Gaussian.
%Our best fit model is close to exponential, and thus has a smaller scale length, which corresponds to a larger extension.
%For our best fit $G_1$, $E_2$ and $E_3$ model, the mean distance from the Galactic center along the major axis is 0.94, 0.66 and 0.43 kpc, the standard deviation of the mean distance is 0.71, 0.66 and 0.52 kpc respectively.

When using all the data, we found an excess, consistent with the so-called X-shape structure in the bulge region. The excess is $\sim$ 5.8\%, approximately in agreement with model of \citet{2012ApJ...757L...7L} when evaluated with their method, which complicated the fitting. It is peculiar that the extrema of the X-shaped structure are at the distance of twice the scale length. Its appearance may be perturbed by disk contamination. As shown in Fig. \ref{residual:E3}, there are fluctuations in the large longitude areas (l $\sim$ 10$^\circ$), and thus may not the true excess shown of the X-shaped structure. We remove those regions and keep the fields with $-4^\circ \leq l \leq 4^\circ$ and $-4^\circ \leq b \leq 4^\circ$ to address this issue and concentrate on the inner region of the bar. As a result, the effect of the fine structure is reduced and we can fit a smoother inner bar structure.

\subsection{Comparison with \citet{2012A&A...538A.106R}}
\begin{table*}
\begin{tabular}{cccccccccccccc}
  \hline
  % after \\: \hline or \cline{col1-col2} \cline{col3-col4} ...
  Model & $R_0$ (kpc) & $x_0$ (kpc) & $y_0$ (kpc) & $z_0$ (kpc) & $\alpha$ (deg)&$\chi^2_1$& $\chi^2_2$&$\chi^2_3$&$\chi^2_4$&$\chi^2/d.o.f$ \\
  \hline
  $\rm G_1$ &8.23 & 1.00  & 0.56& 0.43 & 13.0 &44097&74272&35489&15688&6.27\\
  $\rm E_2$ &8.15 & 0.51  & 0.34 &0.26 & 13.0 &33056&90302&26821&16804&6.17\\
  $\rm E_3$ &8.13 & 0.58  & 0.33 &0.27 & 13.0 &32164&68300&27402&10506&5.12\\
  \hline
\end{tabular}
\caption{\label{tab:alpha} The same as Table \ref{tab:parameter}, but fix the bar angle to be 13$^\circ$.}
\end{table*}

%The bar we found is more prolate than that found by \citet{1995ApJ...445..716D}. Furthermore,
The bar angle we measure is larger than the value found by \citet{2012A&A...538A.106R} ($13^\circ$).  As they pointed out, the bar angle seems to vary with the range of longitude, latitude and population considered, and is found to be between 10 to 45 degrees. The parameterization of \citet{2012A&A...538A.106R} is optimized for values of the bar angle $\alpha$ at the lower end of values found in the literature. The 2MASS data they use have the advantage that it is all-sky, and thus are not limited by spatial coverage. However, 2MASS is also a shallower survey. For most of the bulge, and indeed nearly the entire region we used in our study, 2MASS does not reach the RC population. The upper red giant population has a much broader luminosity range, and a much greater disc contamination fraction, and so we believe RC is a much better tracer of the bulge/bar. Another point is that the study by \citet{2012A&A...538A.106R} removes a lot of data. Their model does not analyze data below the 2MASS completeness limit in J and $K_s$ (see their \S3.2). This implies that they have a variable population under study in every sightline, as the extinction increases, the effective population under study changes a systematic way. In contrast, OGLE-III probes luminosities substantially fainter than the RC over ~98\% of its survey area, so we are not biased in this manner.

We can further test how the bar angle affects the $\chi^2$ in our own models by forcing the bar angle to be 13$^\circ$ as advocated by \citet{2012A&A...538A.106R}, and optimize all other parameters. The results are presented in Table \ref{tab:alpha}. we find the best-fit $\chi^2/d.o.f$ is much larger than that without fixing the bar angle. The scale length of the major axis is 1.00, 0.51 and 0.58 kpc for the $G_1$, $E_2$ and $E_3$ respectively, which is smaller than 1.18, 0.67, 0.66 kpc when the bar angle is included in the fitting, this is reasonable since the length of the bar decreases with the increase of the orientation angle for a fixed observed bar length.

\subsection{Effect of Metallicity}
\label{sec:metallicity}
A vertical metallicity gradient has been measured in the bulge (\citealt{2008A&A...486..177Z, 2013MNRAS.430..836N}). Currently, the metallicty gradient has only been demonstrated for sightlines far from the plane, for which \citet{2013MNRAS.430..836N} find a shallow value of $\sim$0.07 dex degree$^{-1}$. For sightlines closer to the plane, spectroscopic data indicates that the metallicity gradient levels off \citep{2012ApJ...746...59R}. Thus, the metallicity gradient could in principle be a source of systematic error in our investigation, but only for sightlines far from the plane.

We first refer the reader to \citet{2002MNRAS.337..332S}, who used stellar models to demonstrate that metallicity effects on the RC absolute magnitude are substantially smaller in I-band than in either V-band or K-band, which we argue is one of the principle advantages of studying the bulge in optical data from the OGLE-III photometric survey \citep{2011AcA....61...83S}.

We can constrain the impact of population effects as follows. \citet{Nat12} used  models from the BaSTI stellar database \citep{2004ApJ...612..168P} to find that the predicted value of $dM_{\rm I,RC}/d[M/H] = 0.20$ mag dex$^{-1}$, and $d(V-I)_{\rm RC,0}/d[M/H] = 0.27$ mag dex$^{-1}$. Sightlines with a lower metallicity will thus have a slightly underestimated extinction  and a slightly underestimated apparent distance -- the two effects nearly cancel. As the mean extinction law for the bulge is $A_{I}/E(V-I) = 1.22$, the distance error works out to 0.13 mag in distance modulus per dex of metallicity. As the metallicity variation across our fields should be no greater than 0.2 dex, this works out to a potential distance error of 1.0\%. In contrast, the systematic variation in distance due to the Galactic bar's orientation is ~20\%.

The relative impact of population effects on $N_{RC}$ is even smaller. \citet{2001MNRAS.323..109G} compute the predicted number of RC stars in a stellar population as a function of its age and metallicity, which they show in their Figure 1 and summarize in their Table 1. For ages greater than 6 Gyr, the predicted differences for [Fe/H]=-0.3, 0.0, and +0.20 (characteristic metallicities for the bulge) are or order ~3\%. In contrast, the variation in number counts in the fields studied in this work exceeds 600\%, due to the spatial density profile of the bulge. The systematic error is thus smaller than 1 part in 100.

We thus conclude that our study should not be sensitive to population effects. They do exist, but they are not large. Inclusion of these effects would actually be likely to reduce the quality of our analysis, as they are not adequately characterised across the bulge.

\section{Summary and Discussion}
\label{sec:discussion}

We use the RCG population in the OGLE-III database to fit a tri-axial Galactic bar model and to derive much tighter constraints on the bar parameters. When using all the available data, we found there is evidence for an X-shaped structure in the outer part of the bulge region, with observable properties consistent with those of \citet{2012ApJ...757L...7L}. This fine structure complicated the fitting of our analytic models. We removed data from the area in the outer part of the bulge region and retained the fields with $-4^\circ \leq l \leq 4^\circ$ and $-4^\circ \leq b \leq 4^\circ$ to partly remove disk contamination and to concentrate on the inner region of the bar. This reduced the effect of the fine structure and allowed a fit to the resulting smoother inner bar structure. We find that the bar prefers the $E_3$ model and has a bar angle of $29.4^\circ$. We measure the scale length of major axis of 0.67 kpc with an axis ratio of $x_0:y_0:z_0=1.00:0.43:0.40$ for the $E_3$ model. This bar is close to being prolate, and not as triaxial as the bar structure found by \citet{Rat07}; it appears vertically thicker than the working model of \citet{2002ASPC..273...73G} who gives $x_0:y_0:z_0=1.00:0.40:0.30$. For the best fit $E_3$ model, the predicted total number of RCGs for the bulge is 1.05$\times 10^7$. Following the method outlined in \citet{Nat12}, we estimate the total bar stellar mass is 1.8$\times10^{10} M_{\odot}$, about 10\% lower but more accurate than that in \citet{Nat12}.

The model derived here can effectively reproduce the number density, distance modulus over the sky probed by the OGLE-III data (\citealt{Nat12}). Future microlensing observations will provide us with additional constraints on the structure of the Galactic bar. Based on the improved and tightly constrained bar model, it will be interesting to re-calculate the optical depths and time-scale distribution for microlensing events, and see whether they are consistent with the data.

The X-shaped structure appears in both simulations and in our data. Its existence in the Milky way implies that the Galactic bulge is shaped mainly by internal disk dynamical instabilities instead of mergers (e.g. \citealt{2012ApJ...757L...7L}).  How to model subtle structures such as the X-shaped excess remains a challenging task. One possible way forward is to expand the density distribution into orthogonal basis functions; a similar approach is taken in weak lensing to model complex point spread functions (\citealt{2011MNRAS.417.2465A}); we plan to explore this in the near term.

\section*{Acknowledgments}

We acknowledge the National Astronomical Observatories (NAOC) and Chinese Academy of Sciences (CAS) for financial support(SM). LC acknowledges the support of the NSFC grant (11203028) and the Young Researcher Grant of NAOC. This work is also made possible through the support of a grant from John Templeton Foundation (LC). Andrew Gould acknowledges the NSF grant (AST-1103471) and
is also partly supported by a senior visiting professorship of CAS. We thank Drs. Lia Athanassoula, Junqiang Ge, Ortwin Gerhard and Guoliang Li for helpful discussions and Dr. Radek Poleski for comments.

\bibliographystyle{mn2e}

\label{lastpage}
\newpage

\end{document}